\documentclass[11pt,twoside]{article}
\usepackage{./asp2014}

\aspSuppressVolSlug
\resetcounters

\begin{document}

\title{Fast spectral variations of OBA stars}
\author{A.F.~Kholtygin$^1$, S.~Hubrig$^2$, V.V.~Dushin$^1$, S.~Fabrika$^3$, A.~Valeev$^3$, M.~Schoeller$^4$, A.E.~Kostenkov$^1$ 
\affil{$^1$Astronomy Department, Saint-Petersburg State University, Russia;    \email{afkholtygin@gmail.com}}
\affil{$^2$Leibniz-Institut f\"ur Astrophysik Potsdam (AIP), Potsdam, Germany;  }
\affil{$^3$Special Astrophysical observatory, Nizhnii Arhyz, Russia;            } 
\affil{$^4$European Southern Observatory, Garching, Germany;                    }  
       }

\paperauthor{A.F.~Kholtygin}{afkholtygin@gmail.com}{}{Saint-Petersburg State University}{Aspronomyt}{Saint-Petersburg}{}{198504}{Russia}
\paperauthor{S.~Hubrig}{shubrig@aip.de}{}{Leibniz-Institut fur Astrophysik}{}{Potsdam}{}{14482}{Germany}
\paperauthor{M.~Schoeller}{mschoell@eso.org}{}{European Southern Observatory}{}{Garching}{}{85748}{Germany}

\begin{abstract}
The present work is stimulated by the recent detection of the moderate line profile variations of 
selected lines on a time scale of minutes in FORS 2 spectra of the A0 supergiant HD 92207 \citep{Hubrig-2014b}. 
Recently, we investigated the variability of line profiles of selected OBA stars with the multi-mode focal 
reducer SCORPIO at the 6-meter BTA telescope. We discovered regular variations of H and He lines in the spectra of the 
O-type star HD\,93521 (O9.5III) with periods ~4-5 and ~32-36 minutes. The possible origin of short time-scale spectral 
variations is discussed. 
\end{abstract}

\section{Introduction}

Line profiles in spectra of OBA stars are varying on time scales from days to hours \citep{Kaper-1997,Dushin-2013}. 
The variable bright A0 supergiant HD 92207 has been monitored spectroscopically by  \citet{Kaufer-1997}. 
The authors revealed the presence of non-radial 
pulsations (NRPs) in this star with a period of 27 days. The short time-scale line profile changes in 
the spectra of HD~92207 were detected by~\citet{Hubrig-2014b}, see also \citep{Hubrig-2015,Kholtygin-2015}. 
The authors detected the clearly visible LPVs belonging to different elements in individual subexposures reaching up 
to 3\% in intensities and 
up to 30 km/s in radial velocities. Such short-term periodicity was not known 
before for non-radially pulsating supergiants and its study is crucial for modeling the stellar evolution. 

To test whether short-periodic spectral variations are wide dispersed among OBA stars, we investigated 
the variability of 
line profiles of selected OBA stars with the multi-mode focal reducer SCORPIO at the 6-meter BTA telescope. 
In this paper we present the results of our search for the fast LPVs in the spectra of the O-type star HD\,93521 (O9.5III).

\section{HD 93521: line profile variations}

The spectral observations of the fast rotating star HD\,93521 ($V\sin i = 390\,\mathrm{km/s}$, \citep{Rauw-2008}) 
were carried out in 2015 Jan 19--20 with a multi-mode focal reducer 
SCORPIO mounted in the prime focus of the 6-meter telescope BTA. The SCORPIO is usually used for observations 
of star-like and extended objects \citep{Afanasiev-2005} in integal light with a low spectral resolving power. 
Our observations were made width a slit width of 0.5$^"$ in the spectral range $\lambda\lambda\,4040-5850$ 
with the 3\,s exposure. The spectral resolving power of $R\sim 2000$ and the signal to noise ratio $S/N \sim 2000$.

The total time of observations was $T_{\mathrm{full}} = 76$ minutes. The number of obtained spectra was 529, one of 
them appeared to be a bad quality, 
so that finally 528 SCORPIO spectra were used. The reduction of the SCORPIO spectra was done with the MIDAS package using the standard 
procedures. All spectra were normalized to the continuum level using the approach by \citet{Kholtygin-2006}. 

All analyzed lines in the SCORPIO spectra appeared to be variable. In Fig.~\ref{Fig.LPV_Hdelta_Hgamma} we plot the 
LPVs for H$_{\delta}$ and  H$_{\gamma}$ lines.

\articlefigure[width=1.00\textwidth]{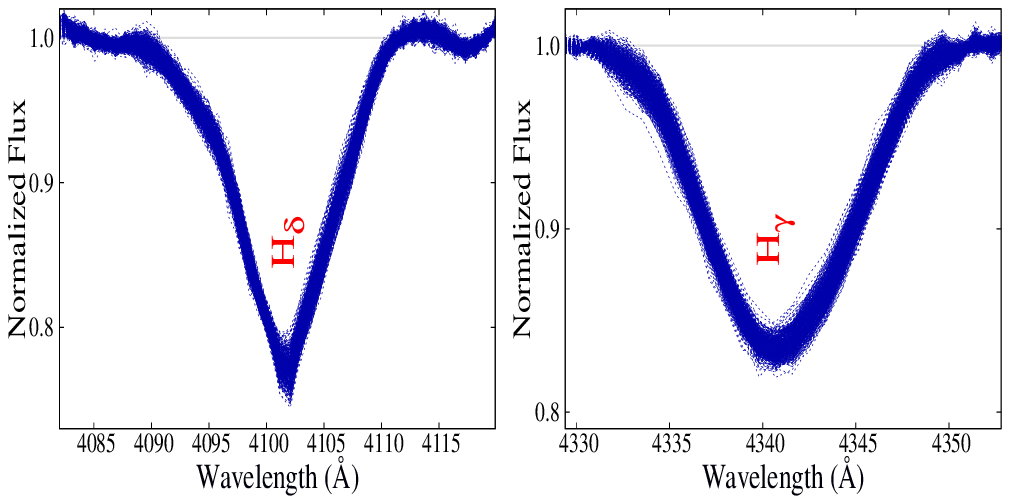}{Fig.LPV_Hdelta_Hgamma} 
   {The overplotted  H$_{\delta}$ and H$_{\gamma}$ line profiles in the SCORPIO spectra  of  HD\,93521. 
    }

%
As it clearly seen in Fig.~\ref{Fig.dfSpMapP}, the line profiles of H and HeII lines change regularly.

\articlefigure[width=1.00\textwidth]{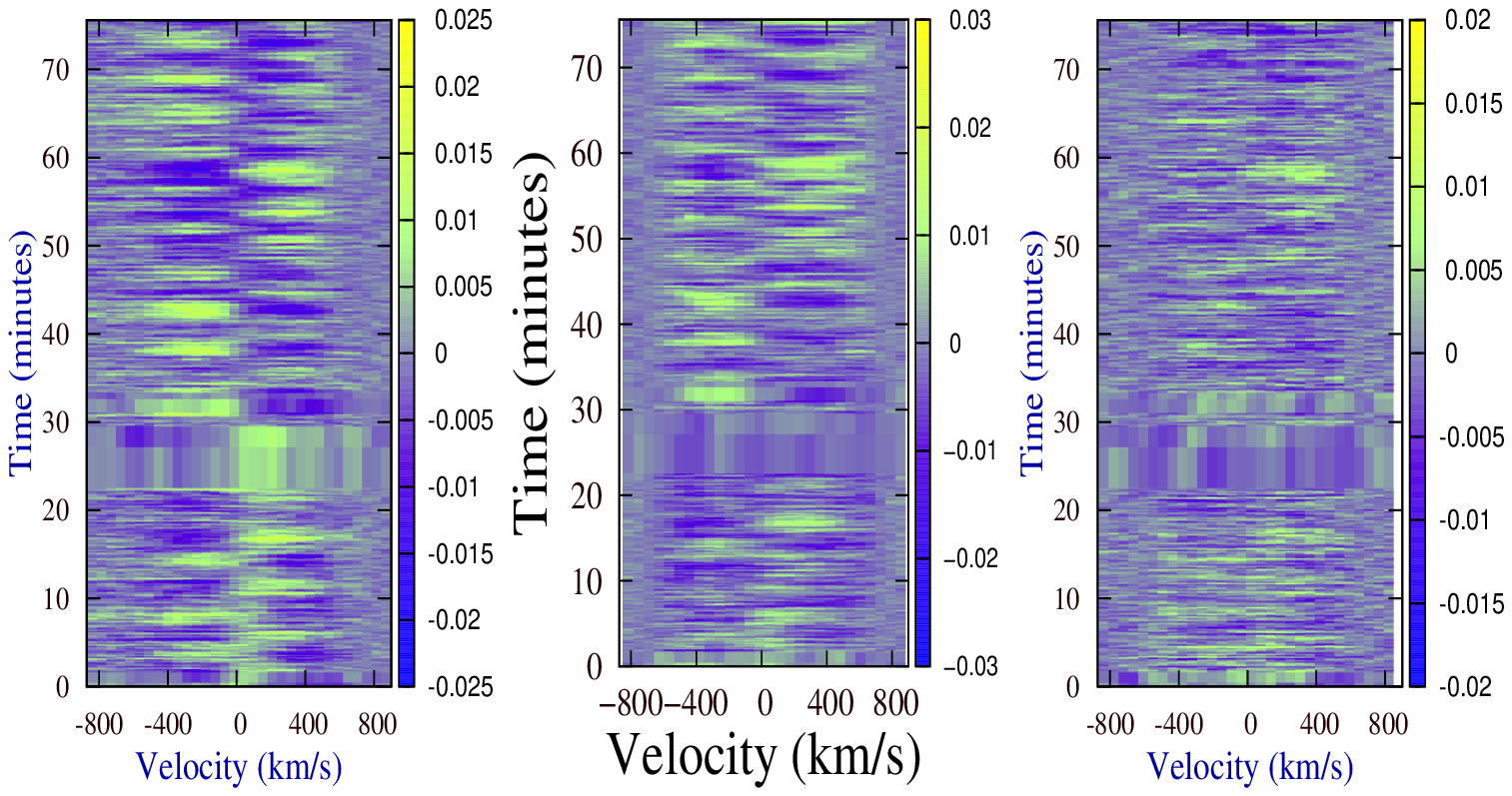}{Fig.dfSpMapP} 
   {The  dynamical spectra of  H$_{\beta}$, H$_{\gamma}$ and HeII$\,\lambda\,4686\,$\AA\ line profiles for HD\,93521.
    }

\section{HD 93521: Fourier spectra}

The frequencies and periods of the LPVs were calculated using the CLEAN algorithm \citep{Roberts-1987} with modification 
by~\cite{Vityazev-1996}. The Fourier spectra of LPVs for H and HeII lines are presented in Fig.~\ref{Fig.Fourier_H-HeP}.

\articlefigure[width=1.00\textwidth]{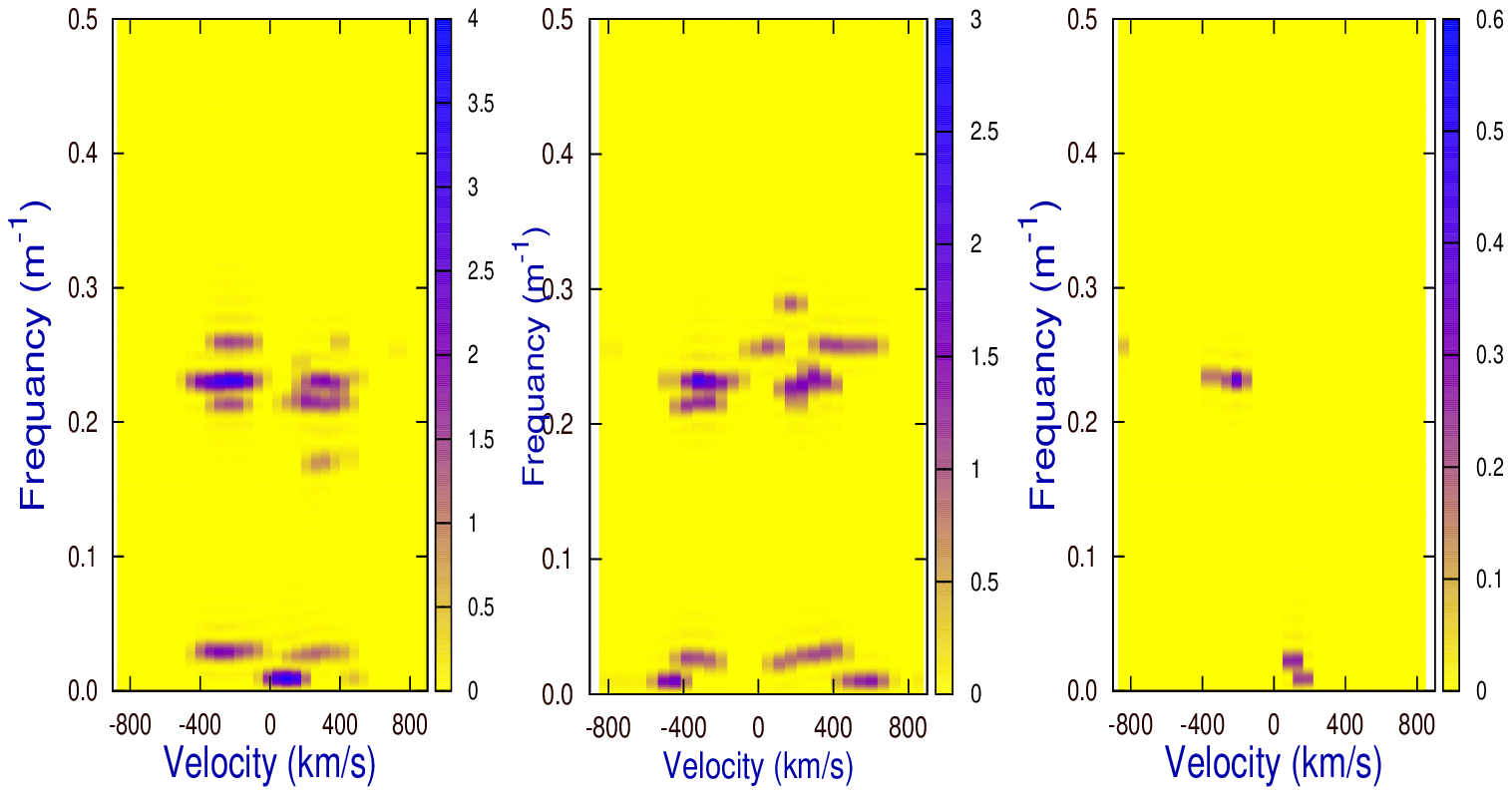}{Fig.Fourier_H-HeP} 
   {The  Cleaned Fourier  spectra for  H$_{\beta}$, H$_{\gamma}$ and HeII$\,\lambda\,4686\,$\AA\ line profiles in spectra HD\,93521.
    }
                 
In Fig.~\ref{Fig.Fourier_H-HeP} one can separate three groups of the regular components in the Fourier spectra: 
$\nu_1=0.21-0.26\,\mathrm{m}^{-1}$ ($P_1=4-5$\,min), $\nu_2=0.028-0.030\,\mathrm{m}^{-1}$ ($P_2=32-36$\,min), and 
$\nu_3 \approx 0.0092\,\mathrm{m}^{-1}$ ($P_3\approx 108$\,min). 

The third component $P_3 >T_{\mathrm{full}}$. In this case, we can not decide whether this component is real. 
However as it was shown by~\cite{Kholtygin-2007a} even in this case such not very reliable component can be close to 
the real 
harmonic of the Fourier spectra if its period only slightly exceed the full length of the time series. 

\citet{Rauw-2008} detected two regular components of HeI and H$_{\alpha}$ LPVs in the spectra of HD~93521 with 
periods of 1.75 and 2.89\,h. 
The authors identified these components as NRP $l=8\pm 1$ and $l=4\pm 1$ modes, respectively.  
Importantly, the shortest period P=1.75~h (105\,min) is very close to our component $P_3$. We suggest that 
our components $P_1$ and $P_2$ also can be assigned to NRP modes $l\sim 25$ and $l\sim 180$, accordingly.

\section{Discussion and conclusion}

The massive OBA stars are considered as type II supernova progenitors. A careful study of their variability provides
important diagnostic means for internal and external (atmospheric) structure.
Our recent search~\citep{Hubrig-2014b} for magnetic fields in selected OBA stars using FOcal Reducer low dispersion 
Spectrographs FORS 1 and FORS 2 in spectropolarimetric mode revealed the presence of strong line profile 
variations (LPV) in spectra of magnetic OBA stars. 

Simple estimations by \cite{Kholtygin-2015} show that the LPVs of such an amplitude as was detected for HD~92207 can be 
connected with the huge spots with diameter about of $40~R_{\odot}$ or with the large prominences with the mean size about 
a few solar radia. 
Regular LPVs in spectra of HD\,93521 are variable on the time scales of ~4-5 and ~32-36 minutes can be connected 
with the high mode of NPPs. The connection of these LPVs with possible magnetic field of the stur is under question.

The number of our spectrapolyrimetric observations of HF~92207 with the minutes time resolution \citep{Hubrig-2014b}  
is too small to find the regular 
component in the LPVs of this star.  Therefore we can not exclude the possibility that LPVs in the spectrum of HD~92207 are 
also associated with high modes of non-radial pulsations like the 5-minute solar oscillations (see, e.g. \citet{Marmolino-1989}).


\acknowledgements The research was supported by the Russian Scientific Foundation (grant N 14-50-00043) in observations with the 
                  BTA telescope, AFK, VVD and AEK acknowledge RFBR for a support by grant 16-02-00604~A in their anylysis of
                  the line profile variations in spectra of HD~93521.

\bibliography{ref}  

\end{document}